\documentclass[reprint,nofootinbib,amsmath,amssymb,aps,prd]{revtex4-1}
\usepackage{graphicx,dcolumn,bm,hyperref,float}

\newcommand{\beq}{\begin{equation}}
\newcommand{\eeq}{\end{equation}}
\newcommand{\bea}{\begin{eqnarray}}
\newcommand{\eea}{\end{eqnarray}}
\newcommand{\Lex}{\mathcal{L}_{\mbox{\tiny{ex}}}}
\newcommand{\LEM}{\mathcal{L}_{\mbox{\tiny{EM}}}}
\newcommand{\Lint}{\mathcal{L}_{\mbox{\tiny{int}}}}
\newcommand{\rhor}{\rho_r}

\begin{document}
	
\title{Symmetries from Locality. I. Electromagnetism and Charge Conservation}

\author{Mark P.~Hertzberg}	
\email{mark.hertzberg@tufts.edu}
\author{Jacob A.~Litterer}
\email{jacob.litterer@tufts.edu}
\affiliation{Institute of Cosmology, Department of Physics and Astronomy, Tufts University, Medford, MA 02155, USA}

\date{\today}
	
\begin{abstract}
It is well known that a theory of the (i) Lorentz invariant and (ii) locally interacting (iii) two degrees of freedom of a massless spin 1 particle, the photon, leads uniquely to electromagnetism at large distances. In this work, we remove the assumption of (i) Lorentz boost invariance, but we still demand (ii) and (iii). We consider several broad classes of theories of spin 1, which in general explicitly violate Lorentz symmetry. We restrict to the familiar two degrees of freedom of the photon. We find that most theories lead to non-locality and instantaneous signaling at a distance. By demanding a mild form of locality (ii), namely that the tree-level exchange action is manifestly local, we find that the photon must still be sourced by a conserved charge with an associated internal symmetry. This recovers the central features of electromagnetism, although it does not by itself impose Lorentz boost symmetry. The case of gravitation dramatically improves the final conclusion and is reported in detail in our accompanying paper Part 2.
\end{abstract}
	
\maketitle

\section{Introduction} 
The basic structure of the fundamental interactions in nature has a long and ongoing history. The modern understanding is that the detailed structure of all the interactions, including electromagnetism, the strong force, the weak force, and gravitation all arise from the postulates of quantum mechanics and local Lorentz invariance. This occurs by introducing and applying these rules at large distances to a particular set of particles: a single massless spin 1 (photon), 8 massless spin 1 (gluons), 3 massive spin 1 (W/Z bosons), and a massless spin 2 (graviton), respectively (e.g., see \cite{WeinbergTextbook}). 

Note that there is currently no known way of deriving the existence of these particles from any low energy point of view, they are in fact postulates (just as the leptons and quarks must be postulated). It is sometimes suggested that they arise from ``gauge symmetry", but this is not meaningful because any theory can be re-written in a gauge invariant way by use of the Stueckelberg trick. Instead, what one can say is that once these degrees of freedom are postulated, along with the rules of Lorentz invariance and locality, the ramifications are profound and currently match observations in great detail. 

In this paper (Part 1) and the accompanying paper \cite{Accompanying2} (Part 2), we would like to explore to what extent one actually requires the assumption of Lorentz boost invariance in order to derive the basic structure of these interactions. Our focus in this paper will be on the simplest case of spin 1, which is relevant to electromagnetism. Without the assumption of Lorentz symmetry, one might wonder what is the starting point. We shall still assume there is a (possibly preferred) frame that exhibits rotation invariance, along with translation invariance in space and time. In earlier work \cite{Hertzberg:2017nzl}, some of us showed that this is in some tension with locality and this issue will be studied in greater generality here. Other works exploring Lorentz violation include Refs.~\cite{Colladay:1996iz,Colladay:1998fq,Coleman:1998ti,AmelinoCamelia:2000mn,Magueijo:2001cr,Muller:2002uk,Kostelecky:2002hh,Kostelecky:2003fs,Collins:2004bp,Dvali:2005nt,Mattingly:2005re,Horava:2009uw,Hohensee:2012dt,Khoury:2013oqa}. 

By imposing Lorentz invariance, it is well known that there is a unique local theory of massless spin 1 photons, coupled to some matter sector. Consistency with causality and unitarity demands a massless photon propagates precisely 2 degrees of freedom (helicities) and its leading interactions with matter are of a uniquely specified form, up to an overall coupling constant and charge assignments \cite{WeinbergTextbook}. 

In this work, we will examine the consequences of relaxing the assumption of Lorentz boost invariance. Usually it is thought that without this principle the entire structure of electromagnetism is completely changed. However, by demanding a rather mild form of locality, namely that there is no instantaneous signaling, we shall examine which aspects remain. In short, we wish to know how far the principle of no instantaneous signaling, without demanding the full structure of Lorentz symmetry, can take us in building this theory. Of particular interest, we will concentrate on the need or otherwise for charge conservation. By the Noether theorem, this is equivalent to examining the need or otherwise for an internal ($U(1)$) symmetry. 
To proceed we first generalize the standard theory of electromagnetism in such a way as to remove the assumption of Lorentz symmetry, and then determine what is recovered upon requiring the theory be local.
As we will show, by projecting down to 2 degrees of freedom for the photon, for reasons we shall explain, we still find that the basic structure of electromagnetism is needed, including a conserved charge, in order to avoid instantaneous signaling. In Part 2  (\cite{Accompanying2}), we study the analogous case for spin 2, which is relevant to gravitation, and examine the need for the Lorentz boost symmetry itself. 

Our paper is organized as follows:
In Section \ref{EM} we discuss locality in electromagnetism and its possible generalizations.
In Sections \ref{TA}, \ref{TB}, and \ref{TC} we examine three distinct classes of theories that deform electromagnetism.
Finally, in Section \ref{Discuss} we discuss our results.

\section{Electromagnetism and its Generalization}\label{EM}

In order to describe the interactions involving massless photons in a manifestly local way, it is convenient to use the electromagnetic field $A_\mu\equiv(-\phi,A_i)$ as a useful mathematical tool. Assuming Lorentz invariance, the leading interactions are given by the usual Maxwell Lagrangian density in the field representation
\beq
\LEM = - \frac{1}{4} F^{\mu\nu} F_{\mu\nu} + \Lint\label{LEM}
\eeq
where $F_{\mu\nu} =  \partial_\mu A_\nu - \partial_\nu A_\mu $. Leading order interactions require minimal coupling with an interaction term of the form
\beq
\Lint= J^\mu A_\mu=J_i A_i - \rho\, \phi 
\eeq
where $\rho$ is the familiar charge density and $J_i$ is the familiar current density. Consistency demands that the current is conserved $\partial_\mu J^\mu=0$ with an associated conserved charge $Q=\int d^3x\,\rho$ \cite{Weinberg:1964ew,Weinberg:1965rz}.

It is of course well known that this theory is local. This is hidden in most gauges, such as Coulomb gauge, but is manifest in Lorenz gauge where $\square A^\mu =J^\mu$. Then for a pair of charged matter sources that undergo tree-level exchange of photons, we can use this equation of motion to readily obtain the tree-level exchange action $\Lex$. Its value is half of the interaction term and is given by
\beq
\Lex = {J_\mu\over2}{J^\mu\over\square} = {J_i\over2} \frac{ J_i }{\square} - {\rho\over2} \frac{ \rho}{\square} \label{EMLint}
\eeq
This result is clearly gauge invariant and therefore we know it will be reproduced in any gauge, such as Coulomb. The presence of the inverse $\square$ wave operator makes it clear that the dynamics is associated with {\em retarded} effects and therefore avoids instantaneous-action-at-a-distance. This is in contrast to an inverse $\nabla^2$ Laplacian operator that would lead to long-range instantaneous effects, violating locality. In Coulomb gauge, such inverse Laplacian operators appear in the equations of motion, but cancel out in the final result \cite{Brill,Gardiner}.

\subsection*{Generalization}
In this work we will assume a preferred frame that still enjoys rotation invariance and translation invariance, but we will not assume boost invariance. We will focus on the lowest dimension operators allowed in the theory. In particular, we will consider the most general quadratic action for the electromagnetic field $A_\mu=(-\phi,A_i)$. We also allow for leading order (non-derivative) coupling to matter, although we do not assume that the matter exhibits current conservation, as that is something that is usually derived by making use of locality and Lorentz symmetry. One of our primary goals is to determine if current conservation is still required without the assumption of Lorentz invariance.

Firstly, we would like to emphasize again that our focus in this work is to just deform away from the Lorentz invariant case. So we are not intending to radically alter the power counting of operators. This means that even though we don't assume exact Lorentz symmetry, the units are essentially the same with mass dimension: $[\phi] = [A_i] = M,\,[J_i] = M^3,\,[\partial_i] = [\partial_t]=M$. For instance, as long as one assumes that the limiting speed of a particle is of the order of the limiting speed of the photon, this suffices for the usual power counting to organize into a hierarchy of relevant operators in the usual fashion.
	
The most general rotationally and translationally invariant action at leading order is then given by expanding out the familiar terms in the Maxwell action, but then allowing for various prefactor couplings, as follows
\bea
\mathcal{L} = \frac{1}{2} \Big{(} \alpha \, \dot{A}_i \dot{A}_i - \beta \, \partial_i A_j \partial_i A_j + \gamma \, \partial_i \phi \partial_i \phi +  2 \epsilon \, \dot{\phi} \partial_i A_i \nonumber \\
+ \delta \, \partial_i A_i \partial_j A_j - m_A^2 A_i A_i + m_\phi^2 \phi^2 \Big{)} + J_i A_i - \rho\, \phi \,\,\,\, \label{5}
\eea
where $\alpha,\beta,\gamma,\epsilon,\delta$ are parameters that generalize the Maxwell Lagrangian. At dimension 4, one could add other terms, like $F[\psi_m] A_i \partial_i \phi,\,G[\psi_m] A_i A_i,\, H[\psi_m] \phi^2$, where $F[\psi_m],\,G[\psi_m],\,H[\psi_m]$ are some functionals of the matter fields denoted $\psi_m$, e.g., we could have a scalar field $\varphi\in\psi_m$, and consider $\sim\varphi \phi^2$, or $\varphi^2 \phi^2$, etc.  Such terms can be added, however they will not be relevant for our analysis. In particular, we will be interested in tree-level interactions between the matter fields, due to photon exchange, as a basic test of locality. In particular, these other terms would be 3 or 4 point vertices, with 2 of the interacting particles being the photon. This means they do not contribute to the tree-level exchange between a pair of matter sources. So while those terms can be present in theory, they cannot help to make the leading order interaction between sources local. Thus our basic conclusions will be unaffected by the addition of these other sorts of terms.

In eq.$~$(\ref{5}) one could set $\alpha=\gamma=1$, without loss of generality, so as to canonically normalize the leading kinetic term for $A_i$ and $\phi$, but we will keep the factors of $\alpha$ and $\gamma$ general in this analysis. Furthermore, one could work in units in which $\beta=1$, which sets the speed of the photon. Note we have also included mass terms $m_A^2$ and $m_\phi^2$ for the vector and scalar potentials, respectively. So in addition to the strength of coupling (the fine-structure constant), this appears as 7 parameters; but due to the redundancy in $\alpha$, $\beta$, and $\gamma$, there are only $7-3=4$ physical parameters here. The Lorentz symmetry would require the reduction in all these parameters down to just the mass, with the usual convention $\alpha=\beta=\gamma=\delta=\epsilon=1$ and $m_\phi^2 = m_A^2$. 

Variation of the action gives the following classical equations of motion 
\bea
- \rho &=&  \gamma \nabla^2 \phi + \epsilon \, \partial_i \dot{A}_i - m_\phi^2 \phi \\
J_i &=& \alpha  \ddot{A}_i - \beta \, \nabla^2 A_i + \epsilon \, \partial_i \dot{\phi} + \delta \, \partial_i \partial_j A_j + m_A^2 A_i  
\eea
As mentioned, giving up Lorentz symmetry means it is far from obvious whether one still requires a conserved current. We can take a linear combination of first derivatives of the above pair of equations in order to shed light on current conservation. The following linear combination, which we denote $\sigma$, parameterizes the breaking of current conservation
\beq
\sigma \equiv \partial_i J_i + \dot{\rhor} \label{sigma}
\eeq
where $\rhor$ is the (re-scaled) charge density $\rhor\equiv(\epsilon/\gamma)\rho$.
We can use the equations of motion to express the possible violation of current conservation directly in terms of the vector potential $A_i$ and $\rhor$ as
\bea
\sigma
&=& \left[ \alpha - \frac{\epsilon^2}{\gamma} \left( \frac{ \gamma \nabla^2 }{\gamma \nabla^2 - m_\phi^2 }\right) \right] \partial_i \ddot{A}_i \nonumber\\
&+& \left[ \left( \delta -  \beta \right) \nabla^2 + m_A^2 \right] \partial_i A_i - m_\phi^2 \frac{ \dot{\rhor} }{\gamma \nabla^2 - m_\phi^2} \label{gencons}
\eea
In the Lorentz invariant case ($\alpha=\epsilon^2/\gamma$ and $\delta=\beta$) and with a massless photon ($m_A=m_\phi=0$) all terms on the right hand side vanish, giving a conserved current $\sigma=0$.

\subsection*{Degrees of Freedom}
In the usual Lorentz invariant case, we know that the degrees of freedom are determined by locality and whether the photon is taken to be massless or massive, with massless implying 2 degrees of freedom (helicities) and massive implying 3 degrees of freedom (spin polarizations). For the massive case, the third longitudinal polarization plays a unique role and behaves quite differently at high energies. While it is well behaved in the case of the abelian theory (assuming the current is conserved) it is well known that it leads to strong coupling problems at high energies in the non-abelian theory. This requires a UV completion in terms of the massless degrees of freedom anyhow, as is the case in the well known Higgs mechanism (e.g., see \cite{HorejsiBook}). 

In the non-Lorentz invariant case, which is our focus in this work, the mapping between the mass of the photon and its degrees of freedom is less clear. However, we will {\em choose} to still project out the third, longitudinal mode. We motivate this as follows: Suppose we consider the non-abelian theory. We know that if it is Lorentz invariant, then we need to project out this mode to avoid unitarity breakdown at high energies, as mentioned above. Let us therefore suppose that we now break Lorentz symmetry, but we do it relatively weakly. This implies that the previous sharp arguments that there would be unitarity problems can remain valid in this regime to some extent too. While the boundary of what constitutes a ``weak" breaking is imprecise, this does motivate us to project out the longitudinal mode and only focus on the two transverse modes from now on. We further note that all observational data is of course consistent with there being only 2 degrees of freedom of the photon \cite{Tanabashi:2018oca}.

We identify three distinct ways to project out the longitudinal mode $\partial_iA_i$. 
We will discuss each of these in detail in this paper; a very brief summary is:
(A) set $\partial_i A_i = 0$, 
(B) choose $\alpha \gamma = \epsilon^2$ and $m_\phi=0$ so that eq.~(\ref{gencons}) fixes $\partial_i A_i$ to be uniquely determined by the sources, and
(C) set $\partial_i A_i = f$ where $f= f[\psi_m]$ is a scalar function of the matter fields and $m_\phi=0$. 
We now systematically study each of these theories and the corresponding tension with locality.

\section{Theory A: Coulomb Constraint}\label{TA}
In this section we consider a direct approach to having only 2 degrees of freedom. We impose the analogue of Coulomb gauge to directly eliminate the longitudinal mode
\beq
\partial_i A_i = 0 
\eeq
which is a natural choice since it makes rotation symmetry manifest. Note that since there is no gauge redundancy in this presentation, this is a choice of theory, not merely a choice of gauge. This can be implemented by adding a Lagrange multiplier into the above theory $\lambda(\partial_i A_i)^2$, which enforces $\partial_i A_i = 0$ on the equations of motion. Then the $\delta\,\partial_iA_i \partial_jA_j$ term in the action does not contribute and so $\delta$ is no longer a parameter of the theory. Hence the number of remaining physical parameters in this theory is $7-3-1=3$.

The equations of motion are
\begin{align}
- \rho &=  \gamma \nabla^2\phi - m_\phi^2\phi \\
J_i &= \left( \square + m_A^2 \right) A_i + \epsilon \, \partial_i \dot{\phi}  .
\end{align}
where $\square \equiv \alpha \partial_t^2 - \beta \nabla^2$ is the wave operator. By forming the above linear combination of charge and current densities, it is simple to check that this theory demands that current obeys
\beq
\sigma= m_\phi^2{\dot\rhor\over\gamma\nabla^2 - m_\phi^2}
\eeq
which is in fact a non-local expression due to the inverse Laplacian; we will return to this point shortly. This strongly suggests that the corresponding theory will exhibit physical non-locality, and we will indeed see this is so.
In these theories, the Coulomb potential is non-dynamical. It obeys a constraint equation and we can readily solve for it as
\beq
\phi=\frac{- \rho}{\gamma \nabla^2 - m_\phi^2}
\label{phic}\eeq
We emphasize that this is an {\em exact} constraint equation. Its validity does not rely upon building an effective field theory with domain of validity only length scales large compared to $1/m_\phi$. Instead we can use this constraint at both long and short distances.

{\em Enforcing locality}: 
Now our goal is to check on locality in the theory, which is highly non-trivial since $\phi$ is instantaneous and is non-zero at finite distances. 
One basic test of locality is that the tree-level exchange action between matter sources is local. To obtain this we need to solve for the vector potential $A_i$. This means we pick up the {\em particular solution} of the above equations of motion and we ignore any homogeneous solutions, which would be associated with photons propagating from or to infinity. The solution for $A_i$ can be written formally in terms of the inverse wave operator as
\beq
	A_i			= \frac{ J_i + \epsilon\frac{\partial_i \dot{\rho}}{\gamma\nabla^2-m_\phi^2} }{\square + m_A^2 }
\eeq
Then we can determine the interaction Lagrangian only in terms of sources as
\bea
\Lex 
		&=&  {J_i\over2} \frac{J_i}{\square + m_A^2} \nonumber\\
		&+&  {\rho\over2}\! \left[ \frac{ \left( \alpha- \epsilon^2{\nabla^2\over\gamma\nabla^2-m_\phi^2}\right) \ddot{\rho} - \beta \nabla^2 \rho + m_A^2 \rho  }{ \left( \square + m_A^2 \right) (\gamma\nabla^2-m_\phi^2)} \right] \label{Lint1}
\label{lexA}\eea
where we have used the conservation law to replace $\partial_i J_i$ and dropped surface terms. The $J_i J_i$ term is local since it is only convoluted with a Klein-Gordon operator, which has a retarded Green's function. 

On the other hand, there are $\rho \rho$ terms that have the inverse operator $(\gamma\nabla^2-m_\phi^2)$. This is a problem for locality, as we explain: If we have a localized source $\rho_1$ and another localized source $\rho_2$ (let us suppose they are each point particles), then the second line of eq.~(\ref{lexA}) tells us that if we separate them a distance $L \lesssim 1/m_\phi$, a change in 1 will affect 2 immediately, due to the inverse $(\gamma \nabla^2 - m_\phi^2)$ operator. Since we are deforming away from the Lorentz invariant theory in which $m_\phi=0$, we wish to consider relatively small values of $m_\phi$ in this work, so this range $L\lesssim 1/m_\phi$ can be rather large distances of interest. Hence, this describes instantaneous interactions over finite distances.

Since we are imposing locality in this work, we must restrict the parameters of the theory to remove the above phenomenon. Recall that the coefficients are all assumed to be constants in this work, so to avoid over-constraining $\rho$, and leading to a trivial theory, we must have that in the $\rho\rho$ term all the inverse Laplacians cancel out. The only way this can happen is if
\beq
	\alpha \gamma = \epsilon^2,\,\,\,\,\,\,\,\,m_A = 0,\,\,\,\,\,\,\,\,m_\phi=0
\eeq
Then the interaction becomes
\beq
\mathcal{L}_{int} = {J_i\over2} \frac{J_i}{\square} - \frac{\beta}{\gamma} {\rho\over2} \frac{\rho}{\square}
\label{Jcoulomb}
\eeq
The precise form of eq.~(\ref{EMLint}) can be recovered exactly by a simple rescaling $\rho\to\rho\sqrt{\beta/\gamma}$ to remove the $\beta/\gamma$ prefactor in the second term. Thus, in this theory, locality requires that $m_A=m_\phi = 0$ and we appear to recover electromagnetism uniquely. We note, however, that this does not imply Lorentz invariance, since the sources may themselves be non-Lorentz invariant. For example, the sound speed of the matter sector may be different than the speed of light. But nevertheless the basic structure of electromagnetism is recovered.

\section{Theory B: Constraint from Equations of Motion}\label{TB}
In the previous section we cut down to two degrees of freedom by manually imposing that $\partial_i A_i$ vanishes. In this section we arrange the theory so that the equations of motion directly fix $\partial_i A_i$. This is done by imposing that the coefficient of the $\partial_i\ddot{A}_i$ term in eq.~(\ref{gencons}) vanishes, rendering the longitudinal mode non-dynamical. This requires 
\beq
\alpha \gamma = \epsilon^2\,\,\,\,\,\,\mbox{and} \,\,\,\,\,\,\, m_\phi = 0
\eeq
Then eq.~(\ref{gencons}) means that $\partial_iA_i$ obeys the constraint 
\beq
		\partial_i A_i 
		= \frac{ \sigma}{\left( \delta - \beta \right) \nabla^2 + m_A^2 }
\eeq
again cutting down to two degrees of freedom. In this theory there are now $7-3-2=2$ physical parameters. On the other hand, the current is completely unconstrained, so there is significant freedom now allowed in the matter sector.

With this constraint on $\partial_i A_i$ the equations of motion become 
\bea
			-\rho &=& \gamma \nabla^2 \phi + \frac{ \epsilon \dot{\sigma}}{\left( \delta - \beta \right) \nabla^2 + m_A^2 } \\
			J_i &=& \left( \square + m_A^2 \right) A_i + \partial_i \left( \epsilon \dot{\phi} + \frac{\delta\, \sigma}{\left( \delta - \beta \right) \nabla^2 + m_A^2 } \right)\,\,\,\,\,\,
\eea
where as before $\square \equiv \alpha \partial_t^2 - \beta \nabla^2$. The solutions are
\bea
\phi &=& \frac{-1}{\gamma \nabla^2} \left( \rho + \frac{ \epsilon \,\dot{\sigma}}{\left( \delta - \beta \right) \nabla^2 + m_A^2 } \right) \\
A_i&= &\frac{J_i}{\square + m_A^2} \nonumber\\
&+& \frac{ \partial_i}{ \left( \square + m_A^2 \right) \nabla^2} \left( \dot{\rhor} + \frac{ \frac{\epsilon^2}{\gamma} \ddot{\sigma} - \delta \nabla^2 \sigma}{\left( \delta - \beta \right) \nabla^2 + m_A^2 } \right)
\eea
which we use to write the interaction Lagrangian only in terms of the sources $\rho$ and $J_i$ and the nonconservation parameter $\sigma$ as
\bea
\Lex &= &
			 {J_i\over2} \frac{J_i}{\square + m_A^2} +  \frac{\rho}{2\gamma} \left[ \frac{ - \beta \nabla^2 \rho + m_A^2 \rho  }{ \left( \square + m_A^2 \right) \nabla^2} \right] \nonumber\\
			&-& \dot{\rhor} \left( \frac{ \left( \delta - \beta \right) \nabla^2 \sigma + m_A^2 \sigma }{ \left( \square + m_A^2 \right) \nabla^2 \left[ \left( \delta - \beta \right) \nabla^2 + m_A^2 \right] } \right) \nonumber\\
			&-& {\sigma\over2} \left( \frac{ \frac{\epsilon^2}{\gamma} \ddot{\sigma} - \delta \nabla^2 \sigma}{\left( \square + m_A^2 \right) \nabla^2 \left[ \left( \delta - \beta \right) \nabla^2 + m_A^2 \right] } \right)
\eea
using eq.~(\ref{sigma}) to replace $\partial_i J_i = \sigma - \dot{\rhor}$ and dropping surface terms (recall in this section we set $\alpha \gamma = \epsilon^2$, which eliminates several terms). 
	
{\em Enforcing locality}:
As before the $J_i J_i$ term is local, but there are nonlocal terms in $\rho \rho$, $\sigma \sigma$, and $\rho\sigma$ cross term. To avoid over-constraining the sources, the only way to eliminate the nonlocality in $\rho \rho$ is by requiring $m_A = 0$. Furthermore, due to the inverse Laplacian in the $\sigma \sigma$ term, we also need to either (i) set $\sigma = 0$, and then enforcing these conditions the interaction becomes identical to eq.~(\ref{Jcoulomb}), or (ii) set $\sigma\propto\nabla^2f$, where $f$ is some local function, and then the Laplacians get canceled. This leads to eq.~(\ref{Jcoulomb}) plus an additional term $\propto f^2$; we will return to this in the next Section, and then in the discussion we will explain why even this is not actually a real modification of electromagnetism.

\section{Theory C: Generalized constraint}\label{TC}
In the previous theory we chose parameters in a special way such that $\partial_i A_i$ is rendered non-dynamical by the equations of motion. There exists a natural generalization of the form
\beq
\partial_i A_i = f, \,\,\,\,\,\mbox{where} \,\,\,\,\,f = f[\psi_m]
\eeq
is some scalar (under rotations) function of the matter fields $\psi_m$. We do not a priori assume that $f$ have any particular form, so we use the notation $f=f[\psi_m]$ to indicate it is a function of some of the matter fields. Nevertheless, we will later find that self-consistency and locality restrict $f$ to be a particular function of the current $J^\mu$. However, eq.~(\ref{gencons}) will in general make $\partial_i A_i$ dynamical by an equation of the form
	\begin{align}
		\partial_i A_i= \frac{ \sigma }{\tilde{\square}} \label{divA}+\ldots
	\end{align}
	where the contractions in both $\tilde{\square}$ and $\sigma \sim \partial_\mu J^\mu$, complicated by the $\alpha,\mathellipsis,\delta$ we've put into the theory, will include some non-unit coefficients as in eqs.~(\ref{sigma}) and (\ref{gencons}). 
The most general way to consistently set $\partial_i A_i$ to be non-dynamical is if we force the wave operator to cancel between numerator and denominator here; i.e. $ \sigma = \tilde{\square} f$ so that the boxes cancel. 

We impose this condition consistently by identifying $f$ as the longitudinal part of $J^\mu$ as follows. First, we decompose the current into its curl-free and divergence-free parts 
	\begin{align}
	J^\mu = M^{\mu\nu} \partial_\nu f + J_\perp^\mu \label{22}
	\end{align}
where $J_\perp^\mu$ is a conserved current ($\partial_\mu J_\perp^\mu \equiv 0$) and $M^{\mu\nu}$ is a matrix that may involves spatial deriatives. Then (\ref{divA}) can be written 
	\begin{align}
		\partial_i A_i &= \frac{ M^{\mu\nu} \partial_\mu \partial_\nu f}{ \tilde{\square}}+\ldots
	\end{align}
	and we need the coefficients in the contractions to be the same in order for the boxes to cancel out and leave $\partial_i A_i$ non-dynamical. That is, from eq.~(\ref{gencons})
	\begin{align}
		M^{\mu\nu} \partial_\mu \partial_\nu f 
		= \tilde{\square} \partial_i A_i - m_\phi^2 \frac{ \epsilon}{\gamma} \frac{ \dot{\rho} }{\gamma \nabla^2 - m_\phi^2}
	\end{align}
	where
	\beq
	\tilde{\square} = \left[ \alpha - \frac{\epsilon^2}{\gamma} \left( \frac{ \gamma \nabla^2 }{\gamma \nabla^2 - m_\phi^2 }\right) \right] \partial_t^2+\left( \delta -  \beta \right) \nabla^2 + m_A^2
\label{tildesquare}	\eeq
Then in order to fix $\partial_i A_i$ we require $m_\phi = 0$ and $M^{\mu\nu} \partial_\mu \partial_\nu = \tilde{\square}$.  
Since we must set $m_\phi=0$, we are left with $7-3-1=3$ parameters here.

To satisfy $M^{\mu\nu} \partial_\mu \partial_\nu = \tilde{\square}$ we  take $M^{\mu\nu}$ to be
\bea
				M^{0 \nu} &=& \left( \alpha - \frac{\epsilon^2}{\gamma} \right) \delta^{0 \nu} \\
				M^{i j} &=& \left( \delta - \beta + \frac{m_A^2}{\nabla^2} \right) \delta^{ij}
\eea
and use eq.~(\ref{22}) to write $\rho$ and $J_i$ in terms of $f$ and $J_\perp^\mu$. We will do this substitution in the end, but in the interest of clarity we choose to do the calculation in terms of $\rho$ and $J_i$. 
	
Then with the imposed constraint $\partial_i A_i = f$ and $m_\phi=0$ the equations of motion are
	\begin{subequations}
		\begin{align}
			-\rho &= \gamma \nabla^2 \phi + \epsilon \dot{f} \\
			J_i &= \left( \square + m_A^2 \right) A_i + \delta \partial_i f + \epsilon \partial_i \dot{\phi} 
		\end{align}
	\end{subequations}
These have the particular solutions
\bea
			\phi &=& \frac{ - \rho - \epsilon \dot{f}}{\gamma \nabla^2} \\
			A_i	&=& \frac{ J_i + \partial_i \left(  -\delta f + \frac{\epsilon}{\gamma} \frac{ \dot{\rho} + \epsilon \ddot{f}}{\nabla^2} \right) }{\square + m_A^2} 
\eea
where $\square \equiv \alpha \partial_t^2 - \beta \nabla^2$ as before. Similarly to the previous sections, we can then use these to rewrite the interaction Lagrangian in terms of the sources and $f$,
\bea
		\Lex 
		&=& {J_i\over2} \frac{J_i}{\square + m_A^2} +  \frac{\rho}{2\gamma} \left[ \frac{ \left( \alpha- \epsilon^2 / \gamma \right) \ddot{\rho} - \beta \nabla^2 \rho + m_A^2 \rho  }{ \left( \square + m_A^2 \right) \nabla^2} \right]  \nonumber\\
		&-&  \dot{\rhor}\! \left[ \frac{ \tilde{\square}_0f }{ \left( \square + m_A^2 \right) \nabla^2 } \right] 
		- {\tilde{\square}_0 f\over2} \!\left[\frac{ - \delta f + \frac{\epsilon^2}{\gamma} \frac{\ddot{f}}{\nabla^2} }{ \square + m_A^2 } \right] 
 \label{Lint2}
\eea
where $\tilde{\square}_0$ is the linear operator of eq.~(\ref{tildesquare}) with $m_\phi=0$. 
The first line of (\ref{Lint2}) is identical to eq.~(\ref{Lint1}). In the second line there are several additional terms involving $f$. 

{\em Enforcing locality}:
As in the previous sections, the $\rho \rho$ terms require $\alpha \gamma = \epsilon^2$ and $m_A = 0$ to be local. Then the interaction becomes
\bea
		\Lex &=& {J_i\over2} \frac{J_i}{\square} - \frac{\beta}{\gamma} {\rho\over2} \frac{\rho}{\square} \nonumber\\
		&-&(\delta-\beta)\dot\rhor{f\over\square}- {1\over2}\left(\delta - \beta \right) f \left( \frac{ \alpha \ddot{f} - \delta \nabla^2 f }{\square} \right)\,\,\, 
\eea
which is entirely local but allows for additional terms that do not appear in eq.~(\ref{EMLint}).

Now using the fact that $f$ is related to $J^\mu$ by eqs.\,(\ref{22}) and  the locality conditions $\alpha \gamma = \epsilon^2$, $m_A = 0$, we have
\bea
		J_i &=& \left( \delta -\beta \right) \partial_i f + J_i^\perp \label{Jf}\\
		\rho &=& \rho_\perp \label{rhoperp}
\eea
Note that one could invert this to write $f$ explicitly in terms of $J^\mu$ as $f=(\partial_\mu J^\mu)/((\delta-\beta)\nabla^2)$. However, it is also useful to view this the other way around: the theory is effectively specified by some scalar function $f=f[\psi_m]$ and a conserved current $J^\mu_\perp$, which determines the current $J^\mu$ that sources $A_\mu$ by the above pair of equations. 
 
We can then write the interaction as 
\bea
		\Lex &=& {J^\perp_i\over2} \frac{J^\perp_i}{\square} - \frac{\beta}{\gamma} {\rho_\perp\over2} \frac{ \rho_\perp }{\square} - {1\over2}\left( \delta - \beta \right) f^2 
		\eea
Thus after enforcing locality in this case we do not simply recover electromagnetism, but we have an additional term controlled by the function $f$ that parameterizes a particular way in which current is not conserved. 

\section{Discussion}\label{Discuss}

\subsection*{Current Conservation}

One may wonder what is the significance of this final term that allows for the current to no longer be conserved, and is parameterized in terms of the arbitrary scalar function $f$. Let us describe this with an example: Suppose that among the various contributions to the matter there includes some real scalar $\varphi\in\psi_m$, with $f=g\,\varphi^2/2$. Then the photon appears to be coupled to a real scalar with current 
\bea
J^\mu = g\,(\delta-\beta)\delta^{\mu i}\,\varphi\,\partial_i\varphi+J^\mu_\perp
\eea
(where $g$ is some coupling). This is certainly very unusual, since we are used to the idea that real scalars cannot couple to photons at leading order; that instead, scalars must be organized into pairs related by charge conjugation, etc. However, there are 2 important observations we would like to make. 

The first is that even though 
$\partial_\mu J^\mu$ need not vanish, eq.~(\ref{rhoperp}) reveals locality has required that the charge density $\rho$ is simply given by $\rho=\rho_\perp$, with $\partial_\mu J^\mu_\perp=0$. And so we still have that
\beq
Q = \int d^3x\,\rho
\eeq
is a conserved charge. Hence no matter how we modify electromagnetism by giving up boost invariance, locality is still enough to preserve the need for a conserved charge. By the reverse form of the Noether theorem, we then need the underlying theory to carry a physical internal symmetry. Recall that of course the Noether theorem does not rely on Lorentz invariance, but only an action principle, which we are sticking to in this work. We know this must be an abelian symmetry, since there is a single conserved charge present. Hence there must be some $U(1)$. (We emphasize that we are not referring to the (small) ``gauge" part of the $U(1)$, which is a redundancy, but the global sub-group that is generated by the conserved charge.) This can be embodied by complex fields in any concrete choice of the visible matter sector $\psi_v\in\psi_m$ that go into forming $J^\mu_\perp=J^\mu_\perp[\psi_v]$, such as by the introduction of Dirac fields. (In the case of complex scalars, one should view this result as applying at leading order in the coupling $g$, since at second order in the coupling one would need to introduce the quartic term $g^2|\Phi|^2 A^2$, as is well known). This is our primary finding.

The second observation is that this result is valid beyond just the tree-level leading order interaction. To explain this, suppose we consider the following action
\beq
\mathcal{L} = -{1\over4}F_{\mu\nu} F^{\mu\nu}+A_\mu J^\mu_\perp-{1\over2}(\delta-1)f^2\label{EMf}
\eeq
where we have just added to the standard electromagnetic action a completely decoupled sector $\psi_d \in \psi_m$, parameterized by some function $f=f[\psi_d]$. This theory has exactly the form of electromagnetism, plus a decoupled piece, and therefore the field $A_\mu$ now enjoys the usual gauge redundancy $A_\mu\to A_\mu+\partial_\mu\alpha$. We can use this to make the following gauge choice
\beq
\partial_i A_i=f
\eeq
which is precisely of the form imposed in Theory (C), but in this case it is simply allowed by the regular gauge invariance, as opposed to specifying a particular theory. One can then manipulate the above term by writing $-{1\over2}f^2={1\over2}(\partial_iA_i)^2-f(\partial_i A_i)$. Upon insertion into the action of eq.~(\ref{EMf}) we see that we recover exactly the form of Theory (C), with $\alpha=\beta=\gamma=\epsilon=1$, $m_\phi=m_A=0$, general $\delta$, and a current given exactly as in eq.~(\ref{Jf}). This proves that the contribution that provided a non-conserved part of the current density was actually from a decoupled sector $\psi_d$, while only a conserved current can couple to the photon and be part of the visible sector $\psi_v$. Note that any form of $f=f[\psi_d]$ is allowed if it only depends on decoupled degrees of freedom. 

On the other hand, if $f=f[\psi_v]$ is a function of some of the visible sector degrees of freedom, then it cannot be completely arbitrary; it must not spoil the conserved charge. For instance, if $f=f[\psi]$, where $\psi\in\psi_v$ is a Dirac field that carries electric charge, then we can have special choices of $f$, such as $f=g\bar\psi\psi$ which respects the $U(1)$ symmetry and hence is compatible with charge conservation. While arbitrary $f=f[\psi_v]$ is not allowed.

\subsection*{Outlook}
In this work we have considered broad classes of theories of interacting spin 1 particles, which exhibit rotation and translation invariance in a preferred frame, but not Lorentz boost invariance. We have found that by imposing that there is no instantaneous signaling, the theory always collapses to the basic structure of Maxwell's electromagnetism. Charge must still be conserved and in fact any non-conserved parts of the current density were only associated with decoupled sectors.

We note that while this recovers the central striking features of electromagnetism, the requirement of current conservation $\partial_\mu J^\mu_\perp=0$, despite appearances, does {\em not} imply Lorentz symmetry. 
In fact it is very easy to allow many different sound speeds in the matter sector (e.g. \cite{Colladay:1998fq}). 
In accompanying work \cite{Accompanying2} we study gravitation in a similar framework and show that in this case the Lorentz boost symmetry can be emergent. This will generalize the earlier work of Ref.~\cite{Hertzberg:2017nzl}. 

Other possible work is to apply our arguments to multiple self-interacting massless spin 1. It would be interesting to see if the principle of no instantaneous signaling is enough to once again recover Yang-Mills theory, as it does in the 
 Lorentz invariant case (e.g., see \cite{WeinbergTextbook}).

\section*{Acknowledgments}
We would especially like to thank McCullen Sandora for help in this work. We also thank Itamar Allali, Mark Gonzalez, Andi Gray, Mudit Jain, Fabrizio Rompineve, Neil Shah, and Shao-Jiang Wang for discussion.
MPH is supported in part by National Science Foundation grant No.~PHY-1720332.

\end{document}